\documentstyle[11pt,newpasp,twoside,epsf]{article}
\def\edcomment#1{\iffalse\marginpar{\raggedright\sl#1\/}\else\relax\fi}
\reversemarginpar

\begin{document}

\title{Effects of Planetesimal Dynamics on the Formation of Terrestrial Planets}

\author{Roman R. Rafikov}

\affil{Institute for Advanced Study,  Einstein Drive, Princeton, NJ, 08540, USA}

\begin{abstract}
Formation of terrestrial planets by agglomeration of planetesimals
in protoplanetary disks sensitively depends on the velocity evolution
of planetesimals. We describe a novel semi-analytical approach to 
the treatment of planetesimal dynamics incorporating the gravitational
scattering by 
 massive protoplanetary bodies. Using this method 
we confirm that planets grow very slowly in 
the outer Solar System if gravitational scattering is the only 
process determining planetesimal velocities, making it hard for
giant planets to acquire their massive gaseous envelopes within
$\la 10^7$ yr. We put forward several possibilities for 
alleviating this problem. 
\end{abstract}

\section{Introduction}

Current paradigm of planetary origin 
(Ruden 1999) assumes that terrestrial 
planets have formed in protoplanetary nebulae 
out of swarms of planetesimals --- rocky or 
icy bodies with initial sizes of several kilometers. 
The same process is thought to 
account for the growth of solid cores of giant planets in the core
instability scenario which postulates that huge gaseous envelopes 
of gas giants were acquired as a result of instability-driven gas
accretion on preexisting cores made of solids (Mizuno 1980).

Our understanding of planetesimal accretion dates 
back to pioneering works by Safronov (1969) who (1) proposed
to use the methods of kinetic theory for investigating the behavior
of large number of planetesimals, and (2) included planetesimal 
dynamics in the picture of their gravitational agglomeration. 
Gravitational scattering between planetesimals tends to excite
their random motions increasing the velocities with which they 
approach each other. This can have an important effect on their merging
 because of the phenomenon of gravitational focusing ---
an enhancement of collision cross-section of two bodies 
through the deflection of their orbits caused by 
their mutual gravitational interaction. Gravitational focusing
increases collision cross-section by a factor 
$1+v_{esc}^2/v_{rel}^2$ over its geometrical value
$\pi (R_{1}+R_{2})^2$, where $R_{1,2}$ are the physical 
radii of colliding planetesimals, $v_{esc}$ is their mutual escape 
velocity, and $v_{rel}$ is the relative velocity of 
planetesimals at infinity. 

From this formula it can be seen that 
gravitational focusing is important only provided that  
$v_{rel}$ is significantly
below $v_{esc}$. 
Safronov's original assumption (1969) was that the biggest bodies in the 
system would be 
able to quickly increase the velocities of surrounding small-mass 
planetesimals to $v_{esc}$ thus rendering further accretion of 
planetesimals by these massive protoplanets inefficient. 
As a result, typical timescale for forming the Earth at $1$ AU from
the Sun is very long --- about $10^8-10^9$ yr. This timescale 
rapidly increases as one goes further out in the Solar System and
reaches $\sim 10^{11}$ yr at $10$ AU from the Sun (roughly present
location of Saturn). This timescale is in stark contrast with the
age of the Solar System (about $4.5$ Gyr) implying that the 
Safronov's assumption of $v_{rel}\sim v_{esc}$ is faulty.

Wetherill \& Stewart (1989) pointed out that at least initially 
planetesimal velocities in protoplanetary disks are not that
large ($v_{rel}\ll v_{esc}$) and are
moderated by mutual planetesimal scattering rather than by a small number 
of very massive bodies (which contain too little mass). They showed 
that in this case planetesimal accretion by massive bodies 
proceeds in a self-accelerating manner 
when most massive objects exhibit fastest growth;  as a result, a 
{\it single} massive object detaches itself 
from the continuous mass spectrum of planetesimals.
This so called ``runaway'' accretion
allows Moon or Mars sized objects to appear on rather 
short timescale (typically $10^4-10^5$ yr) in the terrestrial zone. 
It also seemed to have rescinded the timescale problem for the 
giant planets by enabling their solid cores to grow within the 
gaseous nebula lifetime of several Myr at $5-10$ AU from the Sun
thus allowing them to accrete gas.

The runaway growth scenario was challenged by Ida \& Makino (1993)
who demonstrated using N-body simulations that massive 
protoplanetary ``embryos''  
are in fact able to {\it locally} couple dynamically to the planetesimal 
disk after reaching some threshold mass. The major results of their study
were that (1) massive embryo can strongly ``heat up'' planetesimal 
velocities within several Hill radii of its orbit
(dynamically ``heated zone''), and (2) embryo 
tends to repel planetesimal orbits away from its own orbit 
thus decreasing the 
surface density of small bodies at its location. The former effect 
decreases the role of gravitational focusing while the latter 
lowers the amount of mass which can be accreted by the massive 
body. Both of them act to reduce the accretion rate of the embryo
and this stops its rapid runaway growth.
This accretion regime was termed ``oligarchic growth'' since in this
picture one embryo would reign inside its own heated zone, while
there can be {\it many} such embryos (and their corresponding heated zones)
growing within the local patch of the disk.

Straightforward N-body simulations are neither very well suited for 
determining the 
threshold mass at which transition from runaway to oligarchic growth
occurs as a function of 
planetesimal disk properties, nor they can follow the evolution
of the system for long enough. Although they can  
treat gravitational interactions between planetesimals and the 
embryo directly, without simplifications, they are too time consuming 
and not very flexible. Thus it is important to come up with alternative 
approaches which would be better suited for treating this important 
problem. 

\section{Planetesimal dynamics in the vicinity of protoplanetary embryo}
\label{sect:sect2}
 
Given the large number of planetesimals present in protoplanetary disks
it is natural to employ the statistical approach in studying
their dynamics. At the same time, presence of 
inhomogeneities in the planetesimal disk induced by embryo's gravity 
calls for inclusion of spatial dimension into consideration,
something which conventional one-zone coagulation simulation are 
lacking. 


\begin{figure}  
\plotfiddle{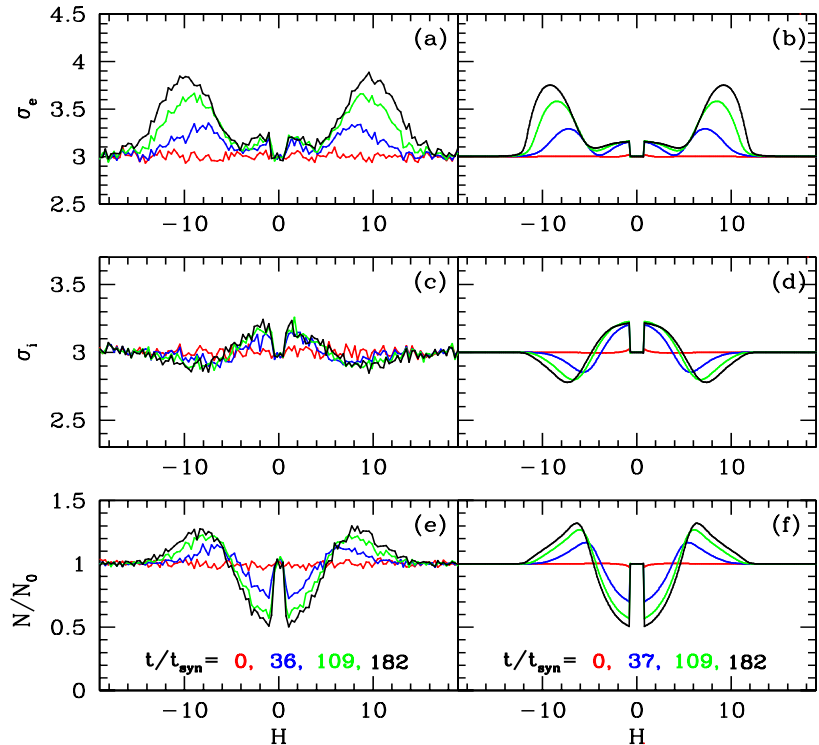}{9.5cm}{0}{60}{55}{-180}{-90}
\caption{Planetesimal disk evolution driven by the 
presence of a protoplanetary embryo. 
The plots contain numerical {\it (left row)} and analytical 
{\it (right row)} time sequences of profiles of $\sigma_e$ 
({\it a,b}), 
$\sigma_i$ ({\it c,d}), and dimensionless surface density 
normalized by
its value at infinity ({\it e,f}). See text for details
[from Rafikov (2003b)].}
\label{fig:fig1}
\end{figure}

Rafikov (2003a, 2003b) came up with an analytical 
statistical prescription for treating planetesimal-planetesimal and
embryo-planetesimal gravitational interactions.
He assumed that the distribution function of planetesimal velocities in
the disk has a Schwarzschild form, which allows one to considerably 
simplify the collision operator in the Boltzmann equation
(describing the evolution caused by 
gravitational perturbations). In this 
approximation, planetesimal disk is unambiguously described by just 
three quantities for each mass population 
--- surface number density $N$, dispersion of
eccentricities $\sigma_e$, and dispersion of inclinations $\sigma_i$
--- which are the functions of radial coordinate in the disk
(all these quantities are assumed to be azimuthally averaged).
A set of three integro-differential equations self-consistently describes
the evolution of these quantities in time and space as mutual 
planetesimal perturbations and gravitational scattering by massive 
protoplanetary embryos cause them to evolve. 

Equations significantly simplify in two regimes:
(1) {\it shear-dominated}, when the 
planetesimal random velocities are small 
compared to the differential shear in the disk 
across the corresponding Hill radius, and
(2) {\it dispersion-dominated}, when these velocities are larger than
the shear across Hill radius. First case can only be appropriate for
the embryo-planetesimal interactions. When it is realized, 
scattering of planetesimals has a deterministic character which 
substantially simplifies its treatment. In the dispersion-dominated 
regime, appropriate for planetesimal-planetesimal scattering 
and in most cases for embryo-planetesimal scattering, 
Fokker-Planck expansion can be performed on the collision operator
and evolution equations reduce to a set of 
partial differential equations. 
Intermediate regime, 
when planetesimal random velocities are comparable to the shear 
across the Hill radius can be treated by 
interpolation between the two limiting cases. 

To check the performance of this approach we have compared 
the results of its application to studying the embryo-planetesimal 
scattering in different velocity 
regimes with the outcome of 
numerical orbit integrations (local) of the same problem
(Rafikov 2003b). 
Representative results for the evolution 
of $\sigma_e, \sigma_i$ and $N$  
in the dispersion-dominated case
(initially $\sigma_e=\sigma_i=3$ in Hill units) 
are shown in Figure 1. Curves of different colors
represent different moments of time (displayed 
in panels ({\it e}) and ({\it f}) by corresponding color coding).
One can see that analytical theory (right panels) is in excellent 
agreement with numerical simulations (left panels) --- they follow 
each other with considerable quantitative accuracy even in minor 
details. This makes one confident that semi-analytical approach
of Rafikov (2003a, 2003b) is quite robust. It is numerically 
inexpensive: analytical calculations displayed in Figure 1 took 
about 1 minute to run on a conventional desktop, while the orbit 
integrations (following 
several $10^5$ particles for several hundred approaches 
to the embryo to achieve enough accuracy) required about 1 month on 
the same hardware. Planetesimal-planetesimal scattering was neglected 
in this calculation (reasonable assumption for large embryo masses),
and only a single mass planetesimal population was considered.
One can clearly see both the local excitation of planetesimal 
velocities and the development of a gap in the surface density of
planetesimal orbits near the embryo, in agreement with
previous N-body simulations. 

This semi-analytical approach thus represents powerful and efficient 
tool for studying the dynamics of planetesimal disks. 
It can be easily 
extended to include the planetesimal-planetesimal scattering (see below), 
planetesimal mass spectrum and its evolution, 
dynamics of {\it several} embryos, dissipative effects, etc.

\section{Growth of an isolated embryo}

As a particular application of this approach we have 
studied the growth of an isolated embryo in a single mass 
planetesimal disk evolving dynamically and spatially under the action
of {\it both} embryo-planetesimal and planetesimal-planetesimal 
gravitational interactions (Rafikov 2003c). Planetesimal-planetesimal
scattering acts as an effective viscosity in the planetesimal disk
opposing the embryo's tendency of clearing a gap.
Masses of planetesimals in a disk are 
assumed not to vary, a simplification justified when embryo's 
mass increases faster than planetesimals grow. This would be
the case in the runaway growth regime, as well as in the oligarchic 
case (Ida \& Makino 1993). At the same time, embryo's mass was allowed
to increase by accreting small planetesimals. 
Accretion rate was calculated 
analytically taking into account inhomogeneity of planetesimal disk
near the embryo and has been checked against the numerical 
orbit integrations (Rafikov 2003b).  

\begin{figure}  
\plotfiddle{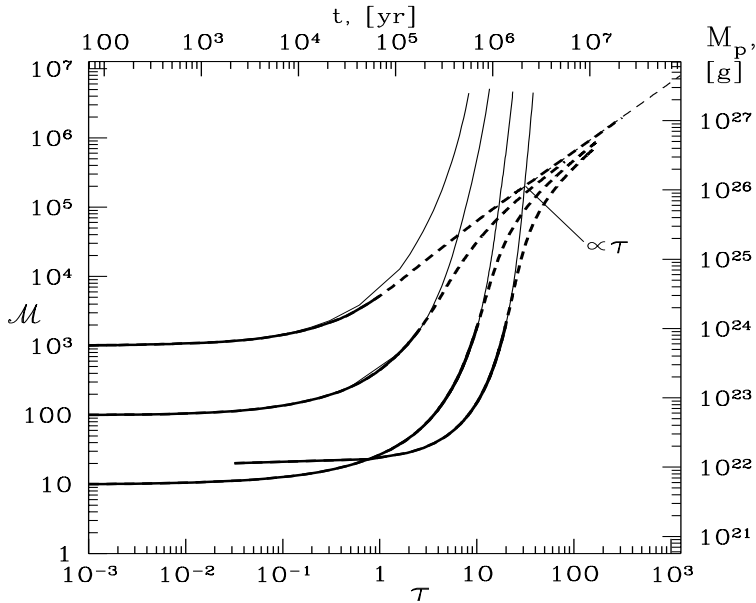}{7.cm}{0}{50}{40}{-150}{-70}
\caption{Growth of the embryo's mass in the Jupiter region
of proto-Solar nebula 
for different initial conditions. See text for details
[from Rafikov (2003c)].} 
\label{fig:fig2}
\end{figure}

Several calculations for different initial starting embryo masses and
planetesimal velocities were carried out and embryo's mass as a 
function of time is displayed in
Figure 2. Properties of the planetesimal disk are those to be expected
at 3.6 AU with planetesimal mass set to $6\times 10^{20}$ g
($\cal M$ is embryo's mass scaled by the planetesimal mass, $\tau$
is time scaled by the synodic period of planetesimals separated by a
Hill radius; corresponding physical values are displayed on the 
right and upper axes). Thin solid lines display the mass evolution 
tracks which embryo follows if its dynamical effect on
its surroundings is neglected. 
As expected, these tracks exhibit unimpeded
runaway growth and embryo reaches very large mass on a rather 
short timescale of $\sim 10^6$ yr. Thick lines represent tracks
with the same initial conditions but with embryo's local 
perturbations taken into account. One can see that they
initially follow the runaway tracks (solid portions). 
This is the result of small embryo's mass, which makes it incapable 
of perturbing planetesimals around it: their velocity dispersions 
increase independently of embryo's growth and 
planetesimal-planetesimal scattering 
is strong enough to smooth any inhomogeneities around the embryo.
However, when embryo grows beyond some threshold mass 
($\sim 10^{24}$ g in this case) it takes
over the control of planetesimal dynamics around it (dashed portions
of thick curves): planetesimals are being pushed away from the embryo's
orbit, clearing a gap (similar to the simple calculation described 
in \S 2), and their velocities increase in accord with the 
embryo's growth.   
As a result, rapid runaway growth changes to a slower power-law increase
of embryo's mass with time, roughly linearly with $t$. Thus, it would 
require a considerably longer timescale (by a factor of $\sim 10$) 
to reach $1 M_\oplus$ than simple runaway picture would predict.
At 5 AU from the Sun this would stretch the 
formation timescale of giant planet cores to $\sim 10^8$ yr which 
is unacceptable from the point of view of core instability scenario 
of giant planet formation given short ($\la 10^7$ yr) lifetimes 
of gaseous nebulae.

\section{Discussion}

Simple problem described above clearly demonstrates 
the difficulty (encountered by conventional scenarios of planet 
formation) of producing solid cores of giant planets 
in the outer Solar System on reasonable timescales. The primary reason 
for this is 
the strong dynamical coupling between massive protoplanetary bodies
and surrounding planetesimals, which causes their gravitational 
focusing to decrease with time making accretion less and less efficient.
In conclusion we want to suggest several possibilities for
curing this problem. 

Embryos likely not have evolved in {\it complete isolation} --- as they
grow in mass their heated zones overlap and they start affecting each 
other's environment. This would likely reduce the tendency for
gap formation around embryo orbits, keeping planetesimal disks 
homogeneous enough to provide the steady supply of planetesimals. 

{\it Dissipative processes} such as 
{\it gas drag} and {\it inelastic collisions} between planetesimals
counteract the tendency of planetesimal velocities to increase
under the action of embryo's perturbations.
And one does expect gas to be naturally present 
during the formation of solid cores of gas giants (and initial
stages of core formation of ice giants). 
This damping would not allow embryos to go back to runaway growth, but 
it would still let them grow faster than if gravity were the only 
force affecting planetesimal velocities.   

{\it Fragmentation} of planetesimals in energetic collisions  
can grind them down to small sizes in the
vicinity of massive bodies. Planetesimals would then be strongly 
affected by dissipative processes and their velocities could 
be considerably reduced allowing embryos to grow faster. 

Closer look at these processes would hopefully help us 
in resolving the issue of planet formation timescale in the 
outer Solar System.

\acknowledgements

The financial support of this work by Charlotte Elizabeth Procter 
Fellowship and W. M. Keck Foundation is gratefully acknowledged.

\end{document}